\documentclass[pra,showpacs,showkeys,nofootinbib,twocolumn]{revtex4}

\usepackage{graphicx}  %
\usepackage{bm}  %

\newcommand{\beq}{\begin{equation}}
\newcommand{\eeq}{\end{equation}}
\newcommand{\bqa}{\begin{eqnarray}}
\newcommand{\eqa}{\end{eqnarray}}

\def\mqo2{{\!\!\!}}

\begin{document}

\title{Enhanced Dimer Relaxation in an Atomic/Molecular BEC}

\author{Eric Braaten}\email{braaten@mps.ohio-state.edu}
\affiliation{Department of Physics,
         The Ohio State University, Columbus, OH\ 43210, USA}

\author{H.-W. Hammer}\email{hammer@phys.washington.edu}
\affiliation{Helmholtz-Institut f{\"u}r Strahlen- und Kernphysik 
   (Abt.~Theorie), Universit{\"a}t Bonn, 53115 Bonn, Germany}
\altaffiliation{Present Address: Institute for Nuclear Theory, 
University of Washington, Seattle, WA 98195, USA}

\date{June, 2004}

\begin{abstract}
We derive a universal formula for the rate constant $\beta$ 
for relaxation of a shallow dimer into
deeply-bound diatomic molecules in the case of
atoms with a large scattering length $a$.
We show that $\beta$ is determined by $a$ 
and by two 3-body parameters that also determine the binding energies 
and widths of Efimov states.
The rate constant $\beta$ scales like $\hbar a/m$ near the resonance,
but the coefficient is a periodic function of $\ln(a)$
that may have resonant enhancement at values of $a$
that differ  by multiples of 22.7. 
\end{abstract}

\smallskip
\maketitle

Using Feshbach resonances \cite{Feshth}, it is
possible to study cold atoms with a variable scattering length $a$
that can be made  large 
enough that the atoms have strong resonant interactions.
Surprising phenomena have been observed in experiments on cold atoms 
with large scattering lengths.
Experiments with a $^{23}$Na BEC \cite{Sten99},
ultracold $^{85}$Rb atoms \cite{Robe00}, a $^{87}$Rb BEC
\cite{Mart02}, and ultracold $^{133}$Cs atoms \cite{Weber03} 
have revealed large losses of atoms as the Feshbach
resonance is approached.
In experiments with a $^{85}$Rb BEC in which $a$ 
was increased briefly to large positive values,
a burst of energetic atoms was observed
along with missing atoms \cite{Cla02}.
More recently, atom-molecule coherence was demonstrated \cite{Don02},
suggesting the coexistence of a condensate of shallow diatomic molecules
with an atom condensate.
Few-body processes play an important role in these phenomena
and it is important to understand their behavior in the limit of large $a$.

A Feshbach resonance allows $|a|$ to be made large compared to the 
typical low-energy length scale $\ell=(mC_6/\hbar^2)^{1/4}$
set by the atom mass $m$ and the van der Waals coefficient $C_6$. 
If $|a|\gg \ell$, the few-body problem
exhibits universal properties that are insensitive to
the details of the interactions responsible for
the large scattering length. 
A simple example in the 2-body sector for $a>0$ is the existence
of a shallow molecule (the {\em dimer}) with binding energy
$|E_2| = \hbar^2/(ma^2)\,$.
A particularly remarkable example in the 3-body sector
is the existence of shallow 3-body bound 
states ({\it Efimov states}) with universal properties \cite{Efi71}.  
As $|a|$ increases, shallower Efimov states
appear below the scattering threshold at values of $a$  that differ
by multiples of $e^{\pi/s_0}\approx 22.7$, where $s_0\approx 1.00624$. 
The Efimov states can cause dramatic dependence 
of scattering observables on $a$ and on the energy.
The 3-body scattering observables
are determined by $a$ and by a 3-body parameter 
that also determines the Efimov spectrum.
Simple universal expressions have been derived 
for the {\it atom-dimer scattering length} \cite{Efi71,BHK99}
and for the rate constant for {\it 3-body recombination} 
into the shallow dimer \cite{NM99,EGB99,BBH00}.  
They are valid for all atoms with $|a|\gg \ell$ 
that have no 2-body bound states if $a<0$ 
and only the shallow dimer if $a>0$.

Unfortunately, heavy alkali atoms, such as Rb and Na, form 
many {\it deeply-bound diatomic molecules}.  
A simple consequence is that Efimov states 
are resonances rather than sharp states, because they
can decay into a deep molecule and an energetic atom.
The widths of Efimov resonances have been studied in Refs.~\cite{NiE00} 
using explicit model potentials.  The deep molecules
also affect other 3-body observables.  In particular,
the universal expressions for the atom-dimer scattering length
and for the 3-body recombination rate constant must be modified.

The existence of deep molecules opens up additional channels 
for the loss of cold atoms from an ultracold gas or a BEC.
One such channel is 3-body recombination into deep molecules,
which has been studied previously in Refs.~\cite{EGB99,BrH01}.
Another loss mechanism for a cold gas containing both atoms and dimers
is {\it dimer relaxation}, in which the collision of
an atom and a shallow dimer
produces a deep molecule and a recoiling atom.  
If the momenta of the incoming atom and dimer are sufficiently small, 
the relaxation rate reduces to a constant.
The {\it relaxation event rate constant} $\beta$ is defined so that the
number of relaxation events per time and per volume in a cold gas 
(or BEC) of atoms and dimers with number densities $n_A$ and $n_D$ 
is $\beta n_A n_D$.
Because of the large energy released,
this process results in the loss of a low-energy atom 
and a shallow dimer:
\beq
\frac{d n_A }{d t} =  \frac{d n_D }{d t}=
 - \beta n_A n_D\,.
\label{eq:defbeta}
\eeq
The loss of atoms from a BEC
through dimer relaxation has been considered previously
in Refs.~\cite{Yur99}. 
The authors assumed that the 
relaxation rate constant $\beta$ is independent of $a$. 
They found qualitative agreement with the 
slow-sweep experiments in \cite{Sten99}
for $\beta\approx 10^{-10}$ cm$^3$/s.

In this paper, we derive a universal expression for the dimer 
relaxation rate constant $\beta$ in the limit of large $a$
using hyperspherical methods combined with previous results
from an  effective field theory.
We show that $\beta$ scales like $\hbar a/m$,
but with a coefficient that is a periodic function of $\ln(a)$
with period $\pi/s_0\approx\ln(22.7)$.
The coefficient of $\hbar a/m$ also depends on 
two 3-body parameters that determine 
the binding energies and widths of the Efimov resonances.
It can have resonant enhancement at those values of $a$ 
for which there is an Efimov resonance near the atom-dimer threshold.

In contrast to other treatments of Feshbach resonances, we do not
attempt to describe the details of the physics in the short-distance region.
Such details include the closed channel of the Feshbach resonance and
the decomposition of the dimer relaxation rate into contributions 
from individual deep 2-body bound states.  
A detailed description of the short-distance physics is not necessary for 
calculating the inclusive dimer relaxation rate.  
Since the total probability is conserved, one can describe the inclusive 
rate by taking advantage of unitarity.  

We first consider the case without deep dimers.
Any probability that flows to the short-distance region where 
the atoms are coupled to the closed channel
must be reflected back to the long-distance 
region in the form of low-momentum states. A detailed knowledge of the 
short-distance dynamics of the Feshbach resonance which creates
the large scattering length is not required. One can eliminate
the closed channel in favor of the scattering length $a$ and the
3-body parameter $\Lambda_*$ discussed below. This simplification
is possible because the resolution scale of the 
low-momentum atoms is given by the scattering length $a$ which is much 
larger than the range of the potential for the closed channel.
As a consequence, low-energy atoms are only able to feel the large
scattering length $a$ (and the three-body parameter $\Lambda_*$
if three- and higher-body observables are concerned) 
but not to probe the mechanism responsible for it.

If deep dimers are present, some of the probability that flows to the 
short-distance region will emerge in the form of high-momentum states 
consisting of an atom and a deep dimer.  
However, if one is interested in the inclusive relaxation rate 
only, it is irrelevant how exactly the probability is lost from the 
low-momentum states.  All one needs to know is how much of 
the probability is lost. In the case of atoms with a large scattering 
length, this can be parametrized by one number $\eta_*$.
The reason for this is again that the resolution scale of the low-momentum
atoms is given by the scattering length $a$ which is much larger than 
the size of the deep dimer.  Therefore the low-momentum atoms 
cannot resolve the details of the physics at short distances comparable
to the size of the deep dimer.

A similar approach was previously used to describe the effect of deep 
2-body bound states on Efimov states \cite{BHK02} or the 
three-body recombination rate \cite{BrH01}.
It has also been applied to many other problems in physics.  
In the mean-field equations for Bose-Einstein condensates, atom loss 
processes are often taken into account through imaginary parts of 
coefficients in the mean-field Hamiltonian \cite{Adhik02}.  
If the loss processes involve momenta large compared to those of atoms 
in the condensate, this prescription can be rigorously justified.  
A similar method is used in nuclear and particle physics to describe 
effects of the annihilation of particles and their antiparticles
\cite{Bodwin:1994jh}.  
The dynamics of the short-distance annihilation process may be very 
complicated, but its effects on low-momentum 
particles can be described by local 
operators in an effective Hamiltonian and all one needs to know is 
the imaginary parts of the coefficients of those terms.

In the remainder of the paper, we will make these arguments more quantitative
and use them to derive a universal expression for the dimer relaxation 
rate constant $\beta$ in the limit of large $a$. We begin by considering
the equation for the hyperradial wavefunction $f(R)$ for the lowest
adiabatic potential in the hyperspherical representation of the 3-body 
problem \cite{Nie01}. 
In the region $\ell \ll R\ll a$, the other adiabatic potentials are
repulsive and the corresponding wavefunctions decrease
exponentially at $R \ll a$. The coupling to those
channels can therefore be neglected and the equation
for $f(R)$ reduces to
\beq
-{\hbar^2 \over 2m} \left[ {\partial^2 \over \partial R^2}
+ {s_0^2 + 1/4 \over R^2} \right] f(R) = E f (R)\,.
\label{radial}
\eeq
This looks like the Schr{\"o}dinger equation for a particle in a
one-dimensional, scale-invariant $1/R^2$ potential.
For low energies $|E| \sim \hbar^2/ma^2$, the energy eigenvalue in 
Eq.~(\ref{radial}) can be neglected and the
most general solution is
\beq
f (R) = \sqrt{H R} \left[A e^{is_0 \ln (H R)} + B e^{-is_0 \ln (H R)}
\right],
\label{f-general}
\eeq
where $H$ has dimensions of wave number but is otherwise arbitrary.
Equation (\ref{f-general}) is simply
the sum of outgoing and incoming hyperradial waves.
At shorter distances $R \sim \ell$ and longer distances $R\sim a$,
the wave function becomes very complicated, but it is constrained
by unitarity.

\begin{figure}[tb]
\centerline{\includegraphics*[width=6cm,angle=0,clip=true]{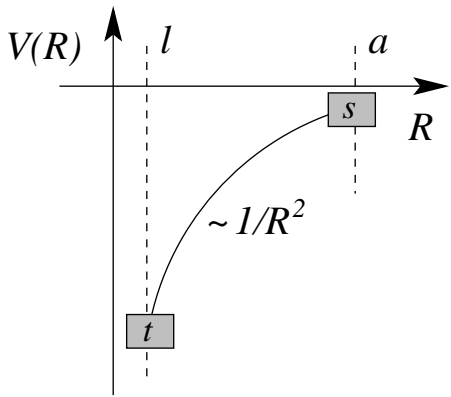}}
\vspace*{0.0cm}
\caption{The hyperspherical potential in the region
 $l \ll R \ll a$. 
The evolution of wavefunctions in the regions $R\sim a$ and $R\sim l$
is described by unitary matrices $s$ and $t$.
}              
\label{fig:1overR2}
\end{figure}

Following Efimov \cite{Efi71}, we first exploit 
unitarity in the long-distance region $R\sim a$.
If we identify the asymptotic states at distances $R \ll a$ 
and $R \gg a$, the evolution of the wave function through 
the region $R \sim a$ can be described by a unitary matrix $s$ 
(see the illustration in Fig.~\ref{fig:1overR2}).
We denote the asymptotic states with probability flowing into 
and out of the region $R \sim a$ by kets $|i \; {\rm in} \rangle$
and $|i \; {\rm out}\rangle$.  
The probability can flow into this region either from
$\ell \ll R \ll a$ or from $R \gg a$.  
In the region $\ell \ll R \ll a$, the asymptotic states 
$|1 \; {\rm in} \rangle$ and $|1 \; {\rm out} \rangle$
are the outgoing and incoming hyperradial waves 
represented by the first and  second terms in Eq.~(\ref{f-general}). 
For simplicity, we restrict ourselves to energies below the dimer-breakup
threshold where 3-atom states are kinematically forbidden.
In the region $R \gg a$, the asymptotic states 
$| 2  \; {\rm in} \rangle$ and $| 2  \; {\rm out} \rangle$
are incoming and outgoing atom-dimer scattering states.
The amplitudes for incoming asymptotic states $| j  \; {\rm in} \rangle$ 
to evolve into outgoing asymptotic states $| i  \; {\rm out} \rangle$
form a unitary $2\times 2$ matrix $s_{ij}$.
Time-reversal invariance implies that $s$ is a symmetric matrix.
Since $s$ is dimensionless, it can depend
on the interaction potential and the energy only through the 
variable $ma^2 E/\hbar^2\,$.

Following Efimov, we next exploit unitarity in
the short-distance region $R\sim \ell$. If there are
no deep molecules, the probability in 
the incoming  hyperradial wave must be totally
reflected from the region $R\sim \ell$ \cite{Efi71}. 
Thus the amplitudes
of the incoming and outgoing hyperradial waves in (\ref{f-general}) 
must be equal in magnitude: $A=-Be^{2i\theta_*}$. The angle  
$\theta_*$ is determined by the logarithmic
derivative $f'(R)/f(R)$ in the region $R \ll a$. This implies that $\theta_*$
must have the form $\theta_*=s_0 \ln({\cal C} a\Lambda_*)$, where
${\cal C}$ is a dimensionless numerical constant and $\Lambda_*$ 
is a parameter that is insensitive to variations in $a$.
The evolution of the wave function in the region $R \sim \ell$
can be described by a $1\times 1$ unitary matrix whose only entry
is $t_{11}=e^{2i\theta_*}$. By combining the constraints from unitarity
in the regions $R\sim \ell$ and $R \sim a$, Efimov obtained an expression 
for the S-matrix element for atom-dimer scattering that he referred to as 
a radial law \cite{Efi71}:
\begin{eqnarray}
S_{AD,AD} & = & s_{22}  + s_{21} 
(1 - t_{11} s_{11})^{-1} t_{11} s_{12}\,.
\label{RLdeep:AD}
\end{eqnarray}
Because $s$ is dimensionless, it can only depend
on the dimensionless variable $ma^2 E/\hbar^2\,$. 
The atom-dimer scattering phase shift can therefore be obtained
by calculating a few universal functions of 
$ma^2 E/\hbar^2\,$ \cite{Efi71}.

We now consider the case with deeply-bound
diatomic molecules.  In this case, some of the probability in 
an incoming hyperradial wave that flows into the region $R \sim \ell$
will emerge as high-energy scattering states of an atom 
and a deep molecule. The amplitude of the 
outgoing hyperradial wave in (\ref{f-general}) must therefore be smaller than 
that of the incoming hyperradial wave: $A=-Be^{2i\theta_*-2\eta_*}$.
The corresponding element of the unitary matrix $t$ is 
\begin{eqnarray}
t_{11} =\exp(2i \theta_* - 2 \eta_*)\,.
\label{t11}
\end{eqnarray}
Inserting this expression into (\ref{RLdeep:AD}), we obtain the 
generalization of Efimov's radial law to the case in which there
are deep molecules. If the universal 
expressions for the S-matrix elements are known in the case $\eta_*=0$,
the effects of deep molecules
can be deduced by the simple substitution 
$\theta_* \to \theta_* + i \eta_*$ \cite{BHK02}.

We proceed to derive the radial law for 
S-matrix elements involving the deep molecules.  
In the short-distance region $R\sim \ell$,
the relevant asymptotic states are incoming and 
outgoing hyperradial waves $|1 \; {\rm in} \rangle$ and
$|1 \; {\rm out}\rangle$ and high-energy atom-molecule scattering
states which we denote by $|X \; {\rm in} \rangle$ and 
$|X \; {\rm out} \rangle$. Note that the outgoing hyperradial wave
is an incoming asymptotic state $|1 \; {\rm in} \rangle$
as far as the region $R\sim a$ is concerned and an outgoing asymptotic state 
$|1 \; {\rm out}\rangle$ as far as the region $R\sim \ell$ is concerned. 
The S-matrix element for a low-energy atom-dimer scattering state to 
evolve into a high-energy atom-molecule scattering state $X$ is
\begin{eqnarray}
S_{X,AD} & = & t_{X1}s_{12}  + t_{X1} s_{11} (1 - t_{11} s_{11})^{-1}
t_{11} s_{12}\,.
\label{RLdeep:X}
\end{eqnarray}
The first term on the right side of Eq.~(\ref{RLdeep:X})
describes transmission through both the regions $\ell \ll R \ll a$
and $R \sim \ell$ to the asymptotic state $X$.
If we expand the second term in powers of $t_{11} s_{11}$, 
we can identify the $n^{\rm th}$ term as the contribution 
from transmission through the region $R \sim a$ 
and reflection from the region $R \sim \ell$,
followed by $n$ reflections from both the regions $R \sim a$ 
and $R \sim \ell$, 
followed by a final reflection from the region $R \sim a$ 
and transmission through the region $R \sim \ell$ 
to the state $X$. 
With each reflection from the region $R \sim \ell$,
the amplitude decreases by a factor $e^{-2\eta_*}$
due to the loss of probability into 
high-energy atom-molecule scattering states.

To calculate the dimer relaxation rate into a specific
high-energy atom-molecule scattering state $X$, we need
the S-matrix element $S_{X,AD}$ in (\ref{RLdeep:X}).
It is extremely sensitive to the details of the interaction 
potential at short distances through the factor $t_{X1}$.
However, the inclusive dimer relaxation rate summed over all 
high-energy atom-molecule scattering states $X$
is much less sensitive to short distances. 
The matrix elements $t_{X1}$ enter only in a combination 
that, by the unitarity of $t$ and the explicit expression for 
$t_{11}$ in (\ref{t11}), depends only on $\eta_*$: 
$\sum_X | t_{X1} |^2 = 1 - e^{-4 \eta_*}\,$.
Squaring the S-matrix element (\ref{RLdeep:X})
and summing over $X$, we obtain
\begin{eqnarray}
{\sum}_X | S_{X,AD} |^2 = (1 - e^{-4 \eta_*}) |s_{12}|^2 
| 1 - t_{11} s_{11} |^{-2}\,.
\label{sumsqSXAD}
\end{eqnarray}
This radial law shows that the inclusive dimer relaxation rate
depends on the interactions only through the parameters $a$, 
$\Lambda_*$, and $\eta_*$. 

To calculate the dimer relaxation rate constant $\beta$,
we need the behavior of the matrix $s$ as $k \to 0$,
where $k$ is the wave number of the atom or dimer in the 
center-of-mass frame. For small $k$, the most general form 
for the entries of $s$ that is allowed by unitarity 
and the analyticity of $S_{AD,AD}$ as a function of $k^2$ is \cite{Efi71}
\begin{eqnarray}
s_{11} &\rightarrow& 
- e^{2 i \phi'} [1 - 2(b_0 + i b_2) ak + \ldots],
\nonumber
\\
s_{12} &\rightarrow& 
e^{i \phi'}\sqrt{4 b_0 ak} [1 - (2 b_3 + i b_1 + i b_2) ak + \ldots],
\nonumber
\\
s_{22} &\rightarrow& 1 - 2(b_0 + i b_1) ak + \ldots,
\label{s-AD}
\end{eqnarray}
where $b_0$, $b_1$, $b_2$, $b_3$, and $\phi'$ are real constants.  
Inserting these expressions for $s_{ij}$ into (\ref{RLdeep:AD}),
and expanding to first order in $k$, we find that the (complex) 
atom-dimer scattering length has the form
\begin{eqnarray}
a_{AD} = (b_1 -b_0  \tan [s_0\ln(a\Lambda_*) + \phi +i\eta_*])a \,.
\label{a12-explicit}
\end{eqnarray}
For $\eta_*=0$, ${\rm Im}\,a_{AD}$ vanishes and Eq.~(\ref{a12-explicit}) 
reduces to Efimov's expression for the atom-dimer scattering length 
in the absence of deep molecules \cite{Efi71}.
The numerical constants $b_1=1.46$, $b_0=2.15$, and $\phi=0.09$ 
were first calculated by Bedaque, Hammer, and van Kolck
using an effective field theory for short-range interactions \cite{BHK99}. 

\begin{figure}[tb]
\centerline{\includegraphics*[width=8.4cm,angle=0,clip=true]{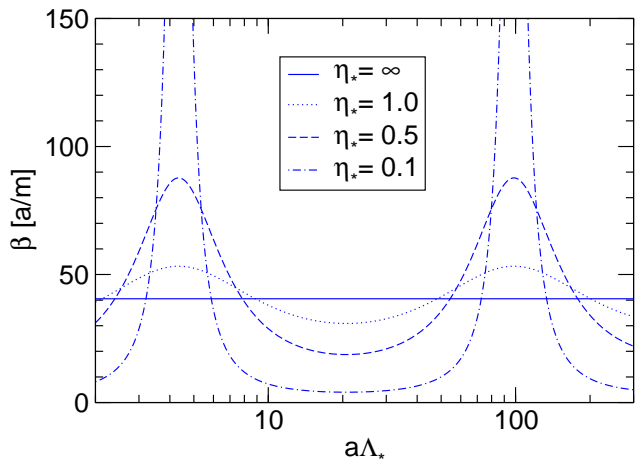}}
\vspace*{0.0cm}
\caption{The dimer relaxation rate constant $\beta$ 
in units of $\hbar a / m$ as a function of $a \Lambda_*$ 
for $\eta_*= 0.1,0.5,1.0$, and $\infty$.}           
\label{fig:ddad}
\end{figure}

The low-energy dimer relaxation rate can be obtained 
by inserting the expansions (\ref{s-AD}) into the radial law
(\ref{sumsqSXAD}).  The  rate constant $\beta$ 
defined in (\ref{eq:defbeta}) is given by the same 
numerical constants $b_0$ and $\phi$ that appear in (\ref{a12-explicit}):
\begin{eqnarray}
 \beta=
{20.3 \;\sinh (2\eta_*) \over \sin^2[s_0\ln(a\Lambda_*) - 1.48] 
 +  \sinh^2 \eta_*}\frac{\hbar a}{m}\,.
\label{beta-ad}
\end{eqnarray}
Alternatively, we can use unitarity to obtain $\beta$ directly from 
(\ref{a12-explicit}): $\beta = - (6 \pi \hbar/m) \, {\rm Im} \, a_{AD}$.
In Fig.~\ref{fig:ddad}, we show the dimer relaxation rate constant $\beta$ 
as a function of $a\Lambda_*$ for a range of $\eta_*$.
For small $\eta_*$, $\beta$ is strongly enhanced
at the values of $a$ corresponding to an Efimov resonance
near the atom-dimer threshold. At the same values of $a$,
there is resonant enhancement of the threshold cross section
$4\pi |a_{AD}|^2$ for atom-dimer scattering. As
$\eta_*$ increases, the resonance structure 
is washed out and the coefficient of $\hbar a/m$ in (\ref{beta-ad})  
approaches a constant: $\beta \approx 40.5 \, \hbar a/m \,$.

For completeness, we also give the universal expressions 
for the 3-body recombination rate constants
when there are deep molecules.
Denoting the event rate constant by $\alpha$, 
the loss rate of low-energy atoms from a cold gas is 
$dn_A/dt=-3 \alpha n_A^3\,$.
In a BEC, the coefficient of $n_A^3$ is smaller by a factor of 6
\cite{Kag85}.
If $a<0$, 3-body recombination can only produce deep molecules,
while if $a>0$ it can also produce the shallow dimer. 
In Ref.~\cite{BBH00}, the rate constant $\alpha$ for $a>0$ was calculated 
as a function of $a$ and $\Lambda_*$ if deep molecules are absent.
The effects of deep molecules on this contribution to $\alpha$
can be deduced using the methods described above.
In Ref.~\cite{BrH01}, the contribution to
$\alpha$ from recombination into deep molecules
was calculated as a function of $a$ and $\Lambda_*$
for both signs of $a$ and infinitesimal $\eta_*$. The
methods described above can be used to generalize those results to
arbitrary $\eta_*$. 
The two contributions to the rate constant for $a>0$ are
\begin{eqnarray}
\alpha_{\rm shallow} &=& 16.8\, {\hbar a^4 \over m}\,
\left(1+e^{-4\eta_*}\right.
\nonumber\\
 & & \left. -2 e^{-2\eta_*}\cos [2s_0 \ln(a\Lambda_*)+0.38]\right)\,,
\label{alpha-sh} 
\\
\alpha_{\rm deep} &=& 16.8 \,{\hbar a^4 \over m}\, \left(1-e^{-4\eta_*} 
\right)\,.
\label{alpha-deep:a>0}
\end{eqnarray}
The only contribution to $\alpha$ for $a<0$ is
\begin{eqnarray}
\alpha_{\rm deep} = {4590 \;\sinh(2\eta_*)\over \sin^2 [s_0 \ln (|a| 
\Lambda_*) -1.38] + \sinh^2 \eta_*}
\; {\hbar a^4 \over m} \;.
\label{alpha-deep:a<0}
\end{eqnarray}
If we set $\eta_* =0$ in (\ref{alpha-sh}), we recover the expression derived 
in Ref.~\cite{BBH00}. If we expand (\ref{alpha-deep:a>0},
\ref {alpha-deep:a<0}) to first order in $\eta_*$, we recover the 
expressions in Ref.~\cite{BrH01} after the identification $\eta_*=
0.2 h_1$ and correcting for a missing factor of $32\pi^2$ 
in Ref.~\cite{BrH01}. The expression in \cite{BrH01} for 
$\alpha_{\rm deep}$ in the case $a>0$ has unphysical
divergences that are artifacts of the expansion to first order in
$\eta_*$. In the general expression (\ref {alpha-deep:a<0}), the divergences
are replaced by resonance peaks.

There have been recent measurements of 3-body recombination rates
for atoms with large scattering length.  In Ref.~\cite{Mart02},
the rate constant $\alpha = {1\over3} K_3$ was measured for 
$^{87}$Rb atoms in the $|1,+1 \rangle$ hyperfine state in the region 
of the Feshbach resonance near 1007 G.  It 
increases by at least two orders of magnitude at the resonance.
Since the off-resonant scattering length is not large, 
the expressions (\ref{alpha-sh})--(\ref{alpha-deep:a<0})
apply only in the resonance region and there were not enough data points 
in this region to observe the scaling behavior $\alpha \sim a^4$.
In Ref.~\cite{Weber03}, the rate constant $\alpha$ was measured for 
$^{133}$Cs atoms in the $|3,+3 \rangle$ hyperfine state in the interval
10 G $<B<$ 150 G, which includes several Feshbach resonances.
Since the off-resonant scattering length for $^{133}$Cs is large, 
the expressions (\ref{alpha-sh})--(\ref{alpha-deep:a<0})
should apply except near the points where $a$ goes through zero.  
The data of Ref.~\cite{Weber03} provides beautiful confirmation of the scaling 
behavior $\alpha \sim a^4$.
No evidence of oscillatory dependence of the coefficient 
of $a^4$ on $\ln(|a|)$ was observed, but neither was it excluded.

Some recent theoretical predictions of 3-body
recombination rates near a Feshbach resonance have given reults inconsistent 
with the scaling behavior $\alpha \sim a^4$.  In Ref.~\cite{YB03},
the scaling behavior $\alpha_{\rm deep}\sim a^2$  was obtained by assuming
a two-step process involving resonant 
production of the shallow dimer followed by dimer relaxation into deep
molecules.  Such a two-step process requires a weak-coupling 
assumption that breaks down in the strong-coupling region where
$a$ is large.  In Ref.~\cite{KM02}, the authors claim that $\alpha$ scales as
$a^2$ for $a>0$ and as $|a|^3$ for $a<0$ 
in a particular model for a Feshbach resonance.    
However their approximations were too crude to extract 
the true asymptotic scaling behavior.

The rate constants $\alpha$ and $\beta$ in
(\ref{beta-ad})--(\ref{alpha-deep:a<0}) 
can be used in mean-field descriptions of condensates 
of atoms and dimers near a Feshbach resonance
to take into account the loss of atoms through 3-body recombination 
and dimer relaxation.
While the scattering length $a$ diverges 
as the magnetic field $B$ passes through the Feshbach resonance,
the 3-body parameters $\Lambda_*$ and $\eta_*$ are smooth functions 
of $B$ that can be approximated by constants if the resonance is 
sufficiently narrow \cite{BrH01}.
The study of loss processes from condensates could be used 
to determine the parameters $\Lambda_*$ and $\eta_*$ for a given
Feshbach resonance.  Once these parameters are known, 
the binding energies and widths of the Efimov states can be predicted
and one can develop a strategy for discovering these remarkable states.  

This research was supported by DOE grant DE-FG02-91-ER4069. 
We thank G.~Shlyapnikov for discussions.

\end{document}